\documentclass[a4paper]{article}
\usepackage{latexsym}
\usepackage{amsthm,amsmath,amssymb}

\usepackage[dvips]{graphicx, color}
\usepackage[title]{appendix}

\usepackage{color}
\usepackage{bbm}
\usepackage{scalerel}
\usepackage{stackengine}

\newtheorem{thm}{Theorem}

\newtheorem*{theorem-non}{Theorem}

\newtheorem{prop}[thm]{Proposition}

\newtheorem*{condition-non}{Condition}

\newtheorem{cor}[thm]{Corollary}

\newtheorem{lem}[thm]{Lemma}

\newcommand{\cE}{{\mathcal E}}

\newcommand{\cM}{{\mathcal M}}

\newcommand{\A}{{\Bbb A}}

\newcommand{\E}{{\Bbb E}}
\newcommand{\F}{{\Bbb F}}

\renewcommand{\O}{{\Bbb O}}

\newcommand{\Z}{{\Bbb Z}}

\def\O{\mathcal O}

\def\lra{\longrightarrow}

\def\={\:=\:}  \def\+{\,+\,}

\def\a{\alpha}   \def\ba{\overline\a}

\def\be{\begin{equation}}   \def\ee{\end{equation}}
\def\bes{\begin{equation*}}   \def\ees{\end{equation*}}
\def\ba{\begin{aligned}}   \def\ea{\end{aligned}}
\def\bc{\begin{cases}}   \def\ec{\end{cases}}
\def\bp{\begin{proof}}   \def\ep{\end{proof}}

\newcommand{\ord}{\mathrm{ord}}

\newcommand{\ov}{\overline}
\def\qqan{\qquad\mathrm{and}\qquad}

\def\smm{\smallsetminus}

\def\la{\lambda}
\def\Ga{\Gamma}

\def\Ker{\mathrm{Ker}}
\def\GL{\mathrm{GL}}

\def\lra{\longrightarrow}

\def\ov{\overline}

\def\l{\ell}

\def\bbm1{\mathbbm 1}

\def\wh{\widehat}
\def\be{\begin{equation}}   \def\ee{\end{equation}}
\def\bes{\begin{equation*}}   \def\ees{\end{equation*}}
\def\bea{\begin{equation}\begin{aligned}}   
\def\eea{\end{aligned}\end{equation}}

\def\mod{\mathrm{mod}\,}

\def\l{\left}

\def\bm{\begin{matrix}}
\def\em{\end{matrix}}
\def\bpm{\begin{pmatrix}}
\def\epm{\end{pmatrix}}
\def\Hom{\mathrm{Hom}}

\def\diag{\mathrm{diag}}

\def\CFm{C_{F, r}(D, g_{I_r}^{~}(mQ))}
\def\Spec{\mathrm{Spec}}
\def\Ext{\mathrm{Ext}}

\def\End{\mathrm{End}}
\def\cEnd{\mathcal{E}\!\mathit{nd}}
\def\cHom{\mathcal{H}\!\mathit{om}}
\def\Pic{\mathrm{Pic}}
\def\cKer{\mathrm{Coker}}

\begin{document}
\title{\bf  Adelic Extension Classes,  
Atiyah Bundles 
and  Non-Commutative  Codes} 
\author{L. Weng}  
\date{}
\maketitle
\begin{abstract} This paper consists of three components. In the first, we give an adelic 
interpretation of the classical extension class associated to extension of locally free 
sheaves on curves. Then, in the second, we use this construction on adelic extension 
classes to  write down explicitly adelic representors in $\GL_r(\A)$ for Atiyah bundles 
$I_r$ on elliptic curves. All these works make sense over any base fields. Finally, as an 
application, for  $m\geq 1$, we construct the global sections of $I_r(mQ)$ in local terms 
and apply it to  obtain rank $r$ MDS codes based on the codes spaces 
$C_{F,r}(D,I_r(mQ))$ introduced in \cite{We}.
\end{abstract}
\section{Locally Free Sheaves in Adelic Language}\label{se1}
Let $X$ be a integral, regular, projective curve of genus $g$ over a finite field $\F_q$.
Denote by $F$  the field of rational functions of $X$, by $\A$ the adelic ring
of $F$, and by $\O$ the integer ring of $\A$.

Let $\cM_{X,r}$ be the moduli stack of locally free sheaves of rank $r$ on $X$. It is well 
known that there is a natural identification among elements of $\cM_{X,r}$ and 
the adelic quotient $\GL_r(F)\backslash \GL_r(\A)/\GL_r(\O)$:
\be
\phi:\, \cM_{X,r}\, \simeq\,\GL_r(F)\backslash \GL_r(\A)/\GL_r(\O).
\ee
Indeed, if $\cE$ is a rank $r$ locally free sheaf on $X$,  the fiber $E_\eta$ of 
$\cE$ over the generic point $\eta$ of $X$ is an  $F$-linear space of dimension $r$.
Among its $\GL_r(F)$-equivalence  class, fix an $F$-linear isomorphism 
\be\label{eq2}
\xi: E_\eta\lra F^r.
\ee
Then there exists a dense open subset $U$ of $X$ such that $\xi$ induces a 
tribulation of $\cE$ on $U$
\be
\xi^{~}_U:\cE\simeq \O_U^{\oplus r}.
\ee

To go local, let $x$ be a closed point of $X$, and denote its formal neighborhood by
$\Spec(\wh\O_x)$, where $\wh\O_x$ is the $x$-adic completion of the local ring 
$\O_{X,x}$. Denote by $\wh F_x$ the fraction field of $\wh O_x$.
Since $X$ is separable, the natural inclusion $F\hookrightarrow \wh F_x$ induces a
morphism 
\be
\wh \iota_x: \Spec(\wh\O_x)\lra X.
\ee
Set $\wh\cE_x:={\wh \iota}^*_x\cE$ the pull-back of $\cE$ on $\Spec(\wh \O_x)$.
Then the fiber $\wh E_{x,\eta}$ of $\wh\cE_x$ over the generic point admits a natural 
$\wh F_x$-linear space of dimension $r$. Furthermore, since $\Spec(\wh\O_x)$ is affine,
for the rank $r$ locally free sheaf $\wh\cE_x$, there exists a free $\wh\O_x$-module 
$ E_x^\sim$ of rank $r$ such that 
\be\label{eq5}
\wh\cE_x\simeq\widetilde{E_x^\sim}\qqan  
E_x^\sim\otimes_{\wh\O_x}\wh F_x=\wh E_{x,\eta},
\ee
where $\widetilde{E_x^\sim} $ denotes the locally free sheaf associated to $E_x^\sim$.
Therefore, induced from the free $\wh\O_x$-module structure on $\E_x$, there is a 
natural isomorphism
\be\label{eq6}
\widetilde \xi_x:\, \wh E_{x,\eta}=E_x^\sim\otimes_{\wh\O_x}\wh F_x\simeq \wh F_x^r.
\ee

On the other hand, induced from $F\hookrightarrow \wh F_x$, there is a natural 
identification $E_\eta\otimes_F\wh F_x=\wh E_{x,\eta}.$ Hence, induced from $\xi$ in  
\eqref{eq2} and the natural inclusion $F\hookrightarrow \wh F_x$, there is  a natural 
isomorphism 
\be\label{eq7}
\wh\xi_x:\,\wh E_{x,\eta}=E_\eta\otimes_F\wh F_x\simeq {\wh F}_x^r.
\ee

Consequently, for the isomorphisms $\widetilde \xi_x$ and $\wh\xi_x$ in \eqref{eq6} and 
\eqref{eq7} respectively, there exists a unique element 
$\wh g_x\in \GL_r(\wh F_x)/\GL_r(\wh \O_x)$ such that the following diagram is 
commutative
\be
\bm
E_\eta\otimes_F\wh F_x&=\wh E_{x,\eta}
=&E_x^\sim\otimes_{\wh\O_x}\wh F_x\\[0.30em]
\wh\xi_x\bigg\downarrow\simeq &&\simeq\bigg\downarrow\, \widetilde\xi_x\\[0.80em]
{\wh F}_x^r&\buildrel{\,g^{~}_{x_{~}}}\over\simeq&{\wh F}_x^r.
\em
\ee
In this way, we obtain an element $(g_x)\in \prod_{x\in X}\GL_x(\wh F_x)$. Furthermore, 
this element $(g_x)$ belongs to $\GL_r(\A)$.
Indeed, by \eqref{eq2}, for $x\in U$, 
\be
\widetilde{E_x^\sim} \simeq {\wh\O}_x^r.
\ee
This implies that, for such $x$, $g_x$ may be choosen to be  the identical matrix 
$I_r:=\diag(1,1,\ldots,1)$. We denote the adelic class of $(g_x)$ constructed above by 
$g_{\cE}$ for later use.

Conversely, for a class in $\GL_r(F)\backslash \GL_r(\A)/\GL_r(\O)$, there exists a 
dense open subset $U$ of $X$ such that $g_x$ are identity matrix for all 
$x\in U$. This yields an trivial locally free sheaf $\O_U^{\oplus r}$  on $U$. On the other 
hand, for a closed point $x$ of $X\smm U$,  the image $g_x^{-1}({\wh\O}_x^{\oplus r})$ 
is a full rank  $\wh\O_x$-module contained in $\wh F_x^r$ since 
$g_x\in \GL_x(\wh F_x)$, and hence induces a locally free sheaf 
$\widetilde{g_x^{-1}({\wh\O}_x^{\oplus r})}$ of rank $r$ on $\Spec(\wh\O_x)$.
All these gives the locally free sheaf $\cE_{(g_x)}$ of rank $r$ on $X$ by the so-called 
fpqc-gluing using the components of $(g_x)$. It is not difficult to see that the adelic 
class $g_{\cE_{(g_x)}}$  coincides with (that of) $(g_x)$.

\section{Extensions of Locally Free Sheaves in Adelic Language}\label{se2}
Over the curve $X$, let 
\be\label{eq10}
\E:\qquad 0\to \cE_1\lra\cE_2\lra\cE_3\to 0
\ee
be a short exact sequence of locally free sheaves.  For $i=1,2,3$, denote by $r_i$ the 
rank of $\cE_i$, and  denote by $g_i\in \GL_{r_i}(\A)$ the adelic class associated to 
$\cE_i$ introduced in \S\ref{se1}. Our aim of this section is to write down $g_2$ in terms 
of $g_1$ and $g_2$ and, more importantly, the extension class associated to $\E$.

\subsection{Classical Approach}
In this subsection, we give a classical description of extension. The main references is 
\cite{H}.

Applying the functor $\Hom_{\O_X}(\cE_3,\cdot)$  to $\E$ leads to a long exact 
sequence of sheaves
\be\label{eq11}
0\!\to \!\Hom_{\O_X}(\cE_3,\cE_1)\!\to\!\Hom_{\O_X}(\cE_3,\cE_2)\!\to\!
\Hom_{\O_X}(\cE_3,\cE_3)\buildrel{\delta}\over \to\!\Ext_{\O_X}^1(\cE_3,\cE_1).
\ee
Following Grothendieck, the extension $\E$, up to  isomorphism, is uniquely determined 
by the $\delta$-image of the identity element $\mathrm{Id}_{\cE_3}$ of 
$\Hom_{\O_X}(\cE_3,\cE_3)$ in $\Ext_{\O_X}^1(\cE_3,\cE_1)$. Being working over the 
curve $X$, we have
\be\label{eq12-}
\Ext_{\O_X}^1(\cE_3,\cE_1)\simeq \Ext_{\O_X}^1(\O_X,\cE_3^\vee\otimes \cE_1)
\simeq H^1(X, \cE_3^\vee\otimes \cE_1),
\ee
where, for a locally free sheaf $\cE$,   we denote its its dual sheaf by $\cE^\vee$.
Note that
\be
\Hom_{\O_X}(\cE_3,\cE_3)\simeq \Hom_{\O_X}(\O_X,\cE_3^\vee\otimes\cE_3)
\simeq H^0(X, \cE_3^\vee\otimes\cE_3),
\ee
\eqref{eq11} is equivalent to the long exact sequence
\be\label{eq14}
0\!\to\! H^0(X,\cE_3^\vee\otimes \cE_1)\!\to\! H^0(X,\cE_3^\vee\otimes \cE_2)\!\to\! 
H^0(X,\cE_3^\vee\otimes \cE_3)\buildrel{\delta}\over \to 
\!H^1(X,\cE_3^\vee\otimes \cE_1).
\ee
In addition, there are natural an  isomorphism and a decomposition
\be
\cE_3^\vee\otimes\cE_3=\cEnd_{\O_X}(\cE_3)\simeq 
\O_X\oplus \cEnd_{\O_X}^0(\cE_3),
\ee
respectively. Here $\cEnd_{\O_X}(\cE_3)$ denotes the sheaf of endmorphisms of 
$\cE_3$ and $\cEnd_{\O_X}^0(\cE_3)$ denotes the sub-sheaf of $\cEnd_{\O_X}(\cE_3)$ 
resulting from the so-called trace zero endmorphisms. Furthermore, since 
$H^0(X, \cEnd_{\O_X}^0(\cE_3))=0$, the morphism $\delta$ in \eqref{eq11} is equivalent 
to the induced morphism
\be
\delta: H^0(X,\O_X)\lra H^1(X, \cE_3^\vee\otimes\cE_1),
\ee
and the extension $\E$, up to  equivalence, is uniquely determined by $\delta(1)$, the 
$\delta$-image of the unit element $1$ of $H^0(X,\O_X)$. Here, as usual, if 
\be
\E':\qquad 0\to \cE_1\lra\cE_2'\lra\cE_3\to 0
\ee
is another  extension of $\cE_3$ by $\cE_1$, and there exists a commutative diagram
\be
\bm
0&\to& \cE_1&\lra&\cE_2&\lra&\cE_3&\to& 0\\
&&\Big\|&&\phi\Big\downarrow \simeq &&\Big\| \\
0&\to& \cE_1&\lra&\cE_2'&\lra&\cE_3&\to& 0\\
\em
\ee
then $\phi$ is called an equivalence  between two extensions  $\E$ and $\E'$ of 
of $\cE_3$ by $\cE_1$. Normally, we denote an equivalence  by $\phi:\E\simeq \E'$.

\subsection{Locally Description}

In this subsection, we give a more concrete local description of the  boundary map
\be
\delta: H^0(X,\cE_3^\vee\otimes \cE_3)\to H^1(X,\cE_3^\vee\otimes \cE_1).
\ee
For this purpose, we first recall the adelic interpretation of 
$H^1(X,\cE_3^\vee\otimes \cE_1)$.  Denote by $g_{\cE}\in \GL_r(\A)$ be an adelic 
element associated to a rank $r$ locally free sheaf $\cE$. Then
\be\label{eq20}
H^1(X,\cE_3^\vee\otimes \cE_1)=\A^{r_1r_3}
\big/(\A^{r_1r_3}(\cE_3^\vee\otimes \cE_1)+F^{r_1r_3}),
\ee
where 
\be
\A^{r_1r_3}(\cE_3^\vee\otimes \cE_1);=\Big\{a\in \A^{r_1r_3}: 
g_{\cE_3^\vee\otimes \cE_1}a\in \O^{r_1r_3}\Big\}.
\ee
It is not difficult to see that this space is isomorphic to
\be\label{eq22}
{\prod}_{x\in X}'\Hom_{\wh F_x}(\wh E_{3;x, \eta}, \wh E_{1,;x,\eta})\Big/
\Big(
\Hom_{\wh \O_x}(\wh E_{3,x}^\sim, \wh E_{1,x}^\sim)+\Hom_{F}(E_{3, \eta},  E_{1,\eta})
\Big).
\ee
Here $\prod'$ denotes the restrict product of 
$\Hom_{\wh F_x}(\wh E_{3;x, \eta}, \wh E_{1,;x,\eta})$ with respect to 
$\Hom_{\wh \O_x}(\wh E_{3,x}^\sim, \wh E_{1,x}^\sim)$. Hence, we should find a 
natural morphism from $\End_{\O_X}(\cE_3, \cE_3)$ to \eqref{eq22} which gives the 
boundary map \eqref{eq11} for the extension classes. 

By applying $\cHom_{\O_X}(\cE_3,\cdot)$, or 
the same $ \cE_3^\vee\otimes$, to the extension $\E$, we obtain a short exact 
sequence of locally free sheaves
\be\label{eq23-}
0\to \cHom_{\O_X}(\cE_3,\cE_1)\to \cHom_{\O_X}(\cE_3,\cE_2)\to 
\cEnd_{\O_X}(\cE_3,\cE_3)\to 0,
\ee
since the  the functor $\cHom_{\O_X}(\cE_3,\cdot)$ and $ \cE_3^\vee\otimes$ 
are left and right exactness, respectively. Furthermore, by applying the derived functor of 
$\Ga(X,\cdot)$ to \eqref{eq23-}, we arrive at  the long exact sequence \eqref{eq11}, 
namely,
\be\label{eq24}
0\!\to \!\Hom_{\O_X}(\cE_3,\cE_1)\!\to\!\Hom_{\O_X}(\cE_3,\cE_2)\!\to\!
\Hom_{\O_X}(\cE_3,\cE_3)\buildrel{\delta}\over \to\!\Ext_{\O_X}^1(\cE_3,\cE_1).
\ee
To understand this boundary map, we next recall that how the boundary map in the  
following well-known snake lemma is constructed.

\begin{lem}[Snake Lemma]
Let $R$ be a commutative ring. Assume that
\be
\bm
0&\to& A_1&\lra& A_2&\lra& A_3&\to& 0\\
&&\phi_1\Big\downarrow\ \ \ &&\ \ \ \Big\downarrow \phi_2&&\ \ \ \Big\downarrow \phi_3&&\\
0&\to& B_1&\lra& B_2&\lra& B_3&\to& 0\\
\em
\ee
is a commutative diagram of $R$-modules with exact rows.
Then, for the kernel and cockerel of $\phi_i$, there is a long exact sequence
\bes
0\!\to\! \Ker(\phi_1)\!\to\!  \Ker(\phi_2)\!\to\!  \Ker(\phi_3)\buildrel\delta\over\lra 
\cKer(\phi_1)\!\to\!  \cKer(\phi_2)\!\to\!  \cKer(\phi_3)\to 0.
\ees
In particular, the boundary mapping $\delta$ is defined by
\be
\bm
\delta:& \Ker(\phi_3)&\lra&\hskip -2.80cm \cKer(\phi_1)\\[0.5em]
&a_3&\mapsto &\phi_2(a_2)\in B_1\ \mod\ \phi_1(A_1),
\em
\ee
where $a_2\in A_2$ is a left of the element $a_3\in A_3$.\footnote{Certainly, $\delta$ is we--defined. Indeed, since $a_3\in \Ker(\phi_3)$ implies that $a_3$ has the zero image  in $cKer(\phi_3)$. This implies that the element $\phi_2(a_2)$ of $\cKer(\phi_2)$ belongs to the sub-module  $\cKer(\phi_1)$.}
\end{lem}

To apply this lemma, we now consider the commutative diagram with exact rows
\be\label{eq25}
\bm
&&0&&0&&0&&\\
&&\big\downarrow&&\big\downarrow&&\big\downarrow&&\\
0&\!\to\!& \Hom_{\wh \O_x}(\wh E_{3,x}^\sim,\wh E_{1,x}^\sim)&\!\to\!&\Hom_{\wh \O_x}(\wh E_{3,x}^\sim,\wh E_{2,x}^\sim)&\to& \End_{\wh \O_x}(\wh E_{3,x}^\sim)&\buildrel\delta\over\to& \\
&&\Big\downarrow&&\Big\downarrow&&\Big\downarrow&&\\
0&\!\to\!& \!\!\!\Hom_{\wh F_x}(\wh E_{3;x,\eta},\wh E_{1;x,\eta})\!\!\!&\!\to\!& 
\!\!\!\Hom_{\wh F_x}(\wh E_{3;x,\eta},\wh E_{2;x,\eta})\!\!\!&\to& 
\!\!\!\End_{\wh F_x}(\wh E_{3;x,\eta})\!\!\!&\to&0\\
&&\Big\downarrow&&\Big\downarrow&&\Big\downarrow&&\\
&\!\buildrel\delta\over\to\!& \Hom(\ov E_{3,x},\ov E_{1,x})&\!\to\!& \Hom(\ov E_{3,x},\ov E_{2,x})&\to& 
\End(\ov E_{3,x})&\to&0\\
&&\big\downarrow&&\big\downarrow&&\big\downarrow&&\\
&&0&&0&&0.&&\\
\em
\ee
Here, for $i=1,2,3$,
\bes
\Hom(\ov E_{i,x},\ov E_{j,x}):=  \Hom_{\wh F_x}(\wh E_{i;x,\eta},\wh E_{j;x,\eta})\big/
\Hom_{\wh \O_x}(\wh E_{i,x}^\sim,\wh E_{j,x}^\sim).
\ees
Obviously, by \eqref{eq22}, we have
\bea\label{eq28-}
&{\prod}_{x\in X}'\Hom(\ov E_{3,x},\ov E_{1,x})\big/\Hom_{F}(E_{3, \eta},  E_{1,\eta})\simeq\\
&{\prod}_{x\in X}'\Hom_{\wh F_x}(\wh E_{3;x, \eta}, \wh E_{1,;x,\eta})\Big/
\Big({\prod}_{x\in X}'
\Hom_{\wh \O_x}(\wh E_{3,x}^\sim, \wh E_{1,x}^\sim)+\Hom_{F}(E_{3, \eta},  E_{1,\eta})
\Big).
\eea

Accordingly, in the senate lemma, we set  $a_2$ to be the identity morphism 
$\mathrm{Id}$ of $\End_{\wh \O_x}(\wh E_{3,x}^\sim)$, by the Snake Lemma, we obtain 
an element $\kappa_x= \delta(\mathrm{Id})$ in  $\Hom(\ov E_{3,x},\ov E_{1,x})$. Therefore, applying the natural quotient morphism
\be
{\prod}_{x\in X}'\Hom(\ov E_{3,x},\ov E_{1,x})\lra
{\prod}_{x\in X}'\Hom(\ov E_{3,x},\ov E_{1,x})\big/\Hom_{F}(E_{3, \eta},  E_{1,\eta})
\ee
we obtain an element $([\kappa_x])\in \Ext_{\O_X}^1(\cE_3,\cE_1)$, which is nothing 
but the extension class for the extension $\E$ of $\cE_3$ by $\cE_1$, by \eqref{eq28-}, 
\eqref{eq22}, \eqref{eq20} and \eqref{eq12-}.  This then completes the proof of the 
following

\begin{thm}\label{t1} The natural bijections
\bes
\bm
\Ext_{\O_X}^1(\cE_3,\cE_1)\simeq H^1(X,  \cE_3^\vee\otimes\cE_1)\\[0.40em]
\Phi\,\bigg\downarrow\simeq\quad\qquad\qquad \simeq\bigg\downarrow \\[0.8em]
{\prod}_{x\in X}'\Hom_{\wh F_x}(\wh E_{3;x, \eta}, \wh E_{1,;x,\eta})\Big/
\Big(
{\prod}_{x\in X}'\Hom_{\wh \O_x}(\wh E_{3,x}^\sim, \wh E_{1,x}^\sim)+\Hom_{F}(E_{3, \eta},  E_{1,\eta})
\Big)
\em
\ees
such that 
\be\label{eq30}
\Phi(\delta(\mathrm{Id}))=\big([\kappa_x] \big)_{x\in X}.
\ee
\end{thm}
\bp
Indeed, the  commutative diagram of bijections are direct consequence of  \eqref{eq22}, 
\eqref{eq20} and \eqref{eq12-}. And the relation \eqref{eq30} comes direct from the 
construction of $\kappa_x$.
\ep
We end this section by an effective construction of the inverse map of $\Phi$.

Let $s_\eta=(s_{x,\eta})$ be be an element of  
$\Hom_{\wh F_x}(\wh E_{3;x, \eta}, \wh E_{1,;x,\eta})$.

To clarifying the structures involved, first we {\it assume that $s_\eta$  is regular for all 
but one closed point $x\in X$.} That is to say, there exists one and only one closed point
$x_0\in X$ such that $s_{x,\eta}$ are regular when $x\not=x_0$.
Choose an  open neighborhood $U_0$ of $x_0$ in $X$ such 
that $s_{x_0,\eta}$ is regular over $U_0\smm\{x_0\}$.
Shrinking $U_0$ if necessary, we may assume that $U_0$ is affine. Denote its affine 
ring by $A_{U_0}$. Within the $r_1+r_3$-dimensional $\wh F_x$-linear space 
$\wh E_{1;x_0,\eta}\oplus \wh E_{3;x_0,\eta}$,
construct an $A_{U_0}$-module  generated by $E_{1,U_0}^\sim \oplus \{0\}$ 
and $\big\{(s_{x_0,\eta}(b),b):b\in E_{3,U_0}^\sim\big\}$, where,  for $i=1,\,3$,  
$E_{i,U_0}^\sim=\Gamma(U_0,\cE_i|_{U_0})$ so that 
$\cE_i|_{U_0}\simeq \widetilde{E_{i,U_0}^\sim}$. In other words, this new 
$A_{U_0}$-module  contained in $E_{1;x,\eta}\oplus E_{3;x,\eta}$  is 
given  by
\be\label{eq28}
E_{1,U_0}^\sim\rtimes_{x_0,s_{x_0,\eta}} E_{3,U_0}^\sim:=\big\{\big(a+s_\eta(b),b\big): 
a\in E_{1,U_0}^\sim,\ b\in E_{3,U_0}^\sim  \big\}.
\ee
Obviously, this $A_{U_0}$-module is free and hence  induces a locally free sheaf which 
we denote by $\cE_1|_{U_0}\rtimes_{x_0,s_{x_0,\eta}} \cE_3|_{U_0}$. That is,
\be
\cE_1|_{U_0}\rtimes_{x_0,s_{x_0,\eta}} \cE_3|_{U_0}
:=\widetilde{E_{1,U_0}^\sim\rtimes_{x_0,s_{x_0,\eta}} E_{3,U_0}^\sim}.
\ee
By our construction, obviously,
\begin{enumerate}
\item [(i)] $\cE_1|_{U_0}$ is a locally free $\O_{U_0}$ sub-sheaf of 
$\cE_1|_{U_0}\rtimes_{x_0,s_{x_0,\eta}} \cE_3|_{U_0}$ such that
\be
\l(\cE_1|_{U_0}\rtimes_{x_0,s_{x_0,\eta}} \cE_3|_{U_0}\right)\big/\l(\cE_1|_{U_0}\right)
\simeq \cE_3|_{U_0}.
\ee
\item[(ii)] Since $s_{\eta}$ is regular over $U\smm\{p\}$, 
\be
\l(\cE_1|_{U_0}\rtimes_{x_0,s_{x_0,\eta}} \cE_3|_{U_0}\right)\big|_{U_0\smm\{x_0\}}
=\cE_1|_{U_0}\oplus \cE_3|_{U_0}.
\ee
\end{enumerate}
In particular, it is possible to glue the locally free sheaves 
$\cE_1|_{U_0}\rtimes_{x_0,s_{x_0,\eta}} \cE_3|_{U_0}$ on $U_0$ and 
$(\cE_1\oplus \cE_3)|_{X\smm\{x_0\}}$ on $X\smm\{x_0\}$ over $U_0\smm\{x_0\}$. 
Denote the resulting locally sheaf by $\cE_1\rtimes_{s_\eta} \cE_3$. Obviously, there is
a natural short exact sequence
\be
\E_{s_\eta}:0\to \cE_1\lra \cE_1\rtimes_{s_\eta} \cE_3\lra\cE_3\to 0.
\ee
Moreover, it is not too difficult to see that $s_\eta$ is equivalent to 
$([\kappa_x])_{x\in X}$ associated to $\E_{s_\eta}$ constructed before 
Theorem\,\ref{t1}.

Now we are ready to treat the general case.   {\it Assume that, as we may, there 
exist closed points $x_1,\ldots, x_n$  such that  
\begin{enumerate}
\item [(1)] for $x\not\in \{x_1,\ldots, x_n\}$,  $s_{x,\eta}$ is regular on $X$, 
\item [(2)] for each $i=1,\ldots, n$, $s_{x_i,\eta}$ is  regular for all but one closed point 
$x_i\in X$
\end{enumerate}
}

\noindent
Similarly as in a single closed point case above,  for each $i$, choose an affine open 
neighborhood $U_i$ of $x_i$ in $X$ such that $U_i\cap U_j=\emptyset$ if $i\not=j$, and 
on each $U_i$, we can construct a locally free sheaf 
$\cE_1|_{U_i}\rtimes_{x_i,s_{x_i,\eta}} \cE_3|_{U_i}$. Consequently, we may glue these 
locally free sheaves on the $U_i$'s with the free sheaf
$(\cE_1\oplus\cE_3)|_{X\smm\{x_1,\ldots,x_n\}}$ on $X\smm\{x_1,\ldots,x_n\}$
on the overlaps $U_i\smm\{x_i\}$. In this way, we obtain a locally free sheaf 
$\cE_1\rtimes_{s_\eta} \cE_3$ of rank $r_1+r_3$ on $X$, 
which is an extension of $\cE_3$ by $\cE_1$. In other words, we obtain a 
short exact sequence of locally free sheaves on $X$
\be
\E_{s_\eta}: 0\to \cE_1\lra \cE_1\rtimes_{s\eta} 
\cE_3\lra\cE_3\to 0.
\ee
In addition, by our construction, it is not difficult to see that  $(s_{i,\eta})$ is equivalent to 
the element $([\kappa_x])$ of $\E_{s_\eta}$constructed before 
Theorem\,\ref{t1}. In this we have constructed the inverse of $\Phi$.
In particular, if all the $s_{x,\eta}$'s are regular, then $s_{x,\eta}$ is equivalent to zero 
and the associated extension is trivial.

\subsection{Adeles and   Locally Free Sheaf from Extension}
As in \eqref{eq10}, let 
\be
\E: 0\to \cE_1\lra \cE_2\lra\cE_3\to 0,
\ee
be a short exact sequence of locally free sheaves on $X$. For each $i=1,2,3$, let 
$g_i=(g_{i,x}^{~})\in \GL_{r_i}(F)\backslash \GL_{r_i}(\A)/\GL_{r_i}(\O)$ be the adelic 
elements associated to $\cE_i$ introduced in \S\,\ref{se1}. 

\begin{prop}\label{p3} Let $([\kappa_x])\in \bigoplus_{x\in X}
\Hom_{\wh F_x}(\wh E_{3;x,\eta},E_{1;x,\eta})
\big/   \Hom_{\wh \O_x}(E_{3,x}^\sim, E_{2,x}^\sim)$ be the extension class of 
Theorem\,\ref{t1}  associated to the extension $\E$. Then, the adelic element
$g_2^{~}=(g_{2,x}^{~})\in \GL_{r_i}(F)\backslash \GL_{r_i}(\A)/\GL_{r_i}(\O)$ of the 
locally free sheaf $\cE_2$ is given by
\be\label{eq35}
g_{2,x}^{~}=\bpm g_{1,x}^{~}&\kappa_x\\[0.5em]
 0^{~}&g_{3,x}^{~}
\epm\qquad\forall x\in X.
\ee
Here $\kappa_x$ are viewed as elements of  the spaces $M_{r_1\times r_3}(\wh F_x)$
of $r_1\times r_3$-matrices with entries in $\wh F_x$.
\end{prop}
\bp This is a direct consequence of the proof of Theorem\,\ref{t1}, particularly, the 
construction of $\Phi^{-1}$ at the end of the previous subsection. Indeed, by 
\eqref{eq28}, we have $g_{2,x}^{~}$ is a upper triangular matrices with diagonal 
blocks $g_{1,x}^{~}$ and $g_{3,x}^{~}$ and with $\kappa_x$ as the right-upper block 
as stated 
in the proposition. Finally, the reason that $([\kappa_x])\in M_{r_1\times r_3}(\A)$ 
comes from the fact that the restrict product $\prod'$ is used in the quotient space 
$\frac{{\prod}_{x\in X}'\Hom_{\wh F_x}(\wh E_{3;x, \eta}, \wh E_{1,;x,\eta})}{
{\prod}_{x\in X}'\Hom_{\wh \O_x}(\wh E_{3,x}^\sim, \wh E_{1,x}^\sim)
+\Hom_{F}(E_{3, \eta},  E_{1,\eta})}$.
\ep

\section{Adelic Interpretations of Atiyah Bundles}
\subsection{Atiyah Bundles over Elliptic Curves}

In the sequel, let $X$ be an elliptic curve over $\F_q$. Then the vector bundle over 
$X$ are classified by Atiyah in \cite{A} using the so-called indecomposable bundles. 
After Mumford introduced the slope stability (\cite{Mu}), it is know that an  
indecomposable bundle over elliptic curves  is semi-stable, and it is stable if and only 
if the rank and the degree of the bundle is mutually prime. (For a proof, see e.g. 
Appendix A of  Tu  \cite{T}.)

\begin{thm}[Atiyah\cite{A}] The semi-stable bundles  on an elliptic curve $X/\F_q$ are 
classified as follows.
\begin{enumerate}
\item [(1)] The bundle $I_n$ defined inductively as the unique non-trivial extension of 
$I_{n-1}$ by $\O_X$. In particular, $I_r^\vee\simeq I_r$.
\item [(2)] Assume that $(r,d)=1$. Then for each line bundle $L$ in the Picard group 
$\Pic^d(X)$ on $X$, parametrizing degree $d$ line bundles on $X$, there exists a 
unique stable bundle $W_r(d;L)$ of rank $r$ and degree $d$ such that
\be
\det W_r(\la)=\la.
\ee
In particular, for any line bundle $L_1\in \Pic^{d_1}(X)$,
\be\label{eq37}
W_r(d;L)\otimes L_1\simeq W_r(d+rd_1;L\otimes L_1^{\otimes r}).
\ee
\item [(3)] Every indecomposable semi-stable vector bundle on $X$ is isomorphic to
$I_r\otimes W_{r'}(\la)$.
\item [(4)] Every semi-stable bundle on $X$ of slope $k/n$ for mutually prime $k$ and 
$n$ is a direct sum of bundles  $I_{nm_1}\otimes W_{n}(L)$ for suitable $m$'s
and $L$'s in $\Pic^{k}(X)$.
Here, as usual, the slope is defined as the degree divided by the rank (of the bundle).
\end{enumerate}
\end{thm}

By \eqref{eq37}, up to tensor by line bundles, it suffices to consider $W_r(d;L)$ such 
that $0<d<r$ and $(r,d)=1$.

In the sequel of the paper, we focus on the Atiyah bundles $I_r$ \ $(r\geq 2)$. The 
case for $W_r(d;L)$ will be discussed elsewhere.

\subsection{Adelic Expressions for   of  Atiyah Bundle $I_r$}
We first give a detailed description for the extension class associated to $I_r$.

\subsubsection{Cases of $I_2$ and $I_3$}

To start with, we consider $I_2$, constructed as the unique non-trivial extension of 
$\O_X$ by 
$\O_X$. Indeed, since $H^1(X,\O_X^\vee \otimes \O_X)\simeq 
H^0(X,\O_X)^\vee\simeq \F_q$,
there exists one and only one non-trivial extension of 
$\O_X$ by $\O_X$, determined by the image of the identity morphism $\mathrm{Id}:\O_X\to\O_X$ in .
\be
{\prod}_{x\in X}'\Hom_{\wh F_x}(\wh F_x,\wh F_x)
\big/\big({\prod}_{x\in X}'\Hom_{ \wh\O_x}(\wh\O_x,\wh\O_x)+\Hom_F(F,F)). 
\ee
Since this space is nothing but
\be
\A/(\O+F)=H^1(X,\O_X)\simeq H^1(X,\O_X^\vee\otimes \O_X)^\vee\simeq \F_q.
\ee
to calculate the extension class $[\kappa_x]$, it suffices to analyze the quotient space 
$\A/(\O+F)$. For this, we fix an $\F_q$-rational point $Q$ of $X$. 

\begin{lem}\label{cl1} The extension class in $\A/(\O+F)$ for $I_2$ is given by
\be
\kappa_{I_2,x}^{~}
=\bc \pi_Q^{-1}&x=Q\\[0.6em]
0&x\not=Q.
\ec
\ee
\end{lem}
\bp
For the rational point $Q$ of $X$, by the vanishing of $H^1(X,\O_X(Q))$, 
\be
\A=\A(Q)+F.
\ee
Therefore, by the first and the second isomorphism theorems
\bea\label{eq45-}
\A/(\O+F)=&(\A(Q)+F)/(\A(0)+F)\simeq\A(Q)/(A(0)+\A(Q)\cap F)\\
&\simeq\frac{\A(Q)/A(0)}{(A(0)+\A(Q)\cap F)/\A(0)}\simeq
\frac{\A(Q)/A(0)}{(A(0)+\A(Q)\cap F)/\A(0)}\\
&\simeq \frac{\A(Q)/A(0)}{(\A(Q)\cap F)/(\A(0)\cap F)}.
\eea
Here, in the last step above, we have 
used  the fact that
\bes
H^0(X,\O_X)=\A(0)\cap F=\A(Q)\cap F=H^0(X,\O_X(Q)).
\ees
Denote by $\pi_Q$ the local parameter of the local field $(\wh F_Q,\wh \O_Q)$ 
associated to $Q$. We have
\be
\A(Q)/\A(0) \simeq (\pi_Q^{-1}\wh\O_Q)/\,\wh\O_Q,
\ee
since
\bes
\A(0)=\prod_{x\in X}\wh\O_x\qqan \A(Q)=\prod_{x\in X\smm\{Q\}}\wh\O_x\times
\big\{a\in\wh F_Q:\pi_Q a\in \wh\O_Q\big\}
\ees
Therefore,
\be
 \A/(\O+F)\simeq  (\pi_Q^{-1}\wh\O_Q)/(\pi_Q^{-1+1}\wh\O_Q)
\simeq (\pi_Q^{-1}\wh\O_Q)/\wh Q_Q\simeq\pi_Q^{-1}\F_q.
\ee
This verifies the assertion in the lemma.
\ep

\begin{cor}\label{c5} An adelic representor $g_{I_2}^{~}=(g_{I_2,x}^{~})$ for the 
Atiyah bundle $I_2$ in $\GL_2(F)\backslash \GL_2(\A)/\GL_2(\O)$ may be chosen as
\be
g_{I_2,x}^{~}=\bc
\bpm 1&\pi_Q^{-1}\\ 0&1
\epm\qquad x=Q\\[1.20em]
\bpm 1&0\\ \ 0\, &\ 1\, 
\epm\qquad x\not=Q.
\ec
\ee
\end{cor}

\bp
This is a direct consequence of Proposition\,\ref{p3} and Lemma\,\ref{cl1}. 
\ep

We end this discussion on $I_2$ by calculating $H^1(X,I_2)$. Induced from the non-
split exact sequence $0\to\O_X\to I_2\to \O_X\to 0$ is a long exact sequence of 
cohomology groups
\bea
0\to& H^0(X,\O_X)\to H^0(X,I_2)\to H^0(X,\O_X)\\
&\to H^1(X,\O_X)\to  H^1(X,I_2)\to H^1(X,\O_X)\to 0
\eea
Since $H^0(X,\O_X)\simeq H^0(X,I_2)\simeq \F_q$ and $H^0(X,\O_X)
\simeq H^1(X,\O_X)$, 
we have  
\bes
H^1(X,I_2)\simeq H^1(X,\O_X)\simeq \F_q.
\ees

\medskip
Next, we consider the Atiyah bundle $I_3$. This is constructed as a non-trivial extension
\be
0\to I_2\lra I_3\lra\O_X\to 0.
\ee
Since $H^1(X,I_2)\simeq\F_q$,  $I_3$ is the unique non-trivial extension of $\O_X$ by 
$I_2$.

To determine the associated extension class $\kappa_{I_3}^{~}$, we first write 
$H^1(F,g^{~}_{I_2})$ as the quotient space $\A^2/(\A^2(g^{~}_{I_2})+F^2)$. Since 
$I_2(Q):=I_2\otimes\O_X(Q)$ is semi-stable of positive degree,  
$H^1(F,g^{~}_{I_2(Q)})=\{0\}$ by the vanishing theorem for semi-stable bundles. This 
implies that
\be
\A^2=\A^2(g^{~}_{I_2(Q)})+F^2
\ee
Thus, similarly to \eqref{eq45-}, by the first and the second isomorphism theorem,
\bea\label{eq52}
H^1(F,g^{~}_{I_2})=&(\A^2(g^{~}_{I_2(Q)})+F^2)\big/(\A^2(g^{~}_{I_2})+F^2)\\
\simeq&\frac{\A^2(g^{~}_{I_2(Q)})\big/\A^2(g^{~}_{I_2})}{(\A^2(g^{~}_{I_2(Q)})
\cap F^2)
\big/(\A^2(g^{~}_{I_2})\cap F^2)}
\eea

\begin{lem} We have
\bea
\A^2(g^{~}_{I_2})=&\prod_{x\in X\smm\{Q\}}\wh\O_x^2\times\big\{ (-\pi_Q^{-1}
a_1+a_2,a_1)\in {\wh F}_Q^2: a_1,a_2\in \wh\O_Q\big\}\\
\A^2(g^{~}_{I_2(Q)})=&\prod_{x\in X\smm\{Q\}}\wh\O_x^2\times\big\{ (-\pi_Q^{-1}
a_1+a_2,a_1)\in {\wh F}_Q^2: a_1,a_2\in \pi_Q^{-1}\wh\O_Q\big\}\\
\A^2(g^{~}_{I_2})\cap F^2=&\big\{(a_2,0)\in F^2:a_2\in \F_q\big\}\simeq \F_q
\times\{0\}\\
\A^2(g^{~}_{I_2(Q)})\cap F^2=&\left(\prod_{x\in X\smm\{Q\}}\wh\O_x^2\times
\left\{\begin{aligned}\pi_Q^{-1}(-\pi_Q^{-1} a_1+a_2,a_1):\\
-\pi_Q^{-1} a_1+a_2=\pi_Qa_2'\\
a_1=\pi_Qa_1',\\  a_1',a_2'\in \wh\O_X\end{aligned}\right\}\right)
\bigcap F^2\simeq\F_q^2
\eea
\end{lem} 
\bp
Our description of the  space $\A^2(g^{~}_{I_2(Q)})$ is a direct consequence for the 
space $\A^2(g^{~}_{I_2})$, so  it suffices to calculate the later.
In addition, by Corollary\,\ref{c5} $g_{I_2}^{~}$ is determined as
\bes
g_{I_2,x}^{~}=\bc\bpm 1&\pi_Q^{-1}\\ 0&1\epm, x=Q\\[1.80em]
\ \ \bpm 1&0\\ 0&1\epm, x\not=Q 
\ec
\ees
we have
\bea
\A^2(g^{~}_{I_2})=&\prod_{x\in X\smm\{Q\}}\wh\O_x^2\times
\big\{ (a_2,a_1)\in {\wh F}_Q^2: \bpm 1&\pi_Q^{-1}\\ 0&1\epm
\bpm a_2\\ a_1\epm\in {\wh\O}^2_Q\big\}\\
=&\prod_{x\in X\smm\{Q\}}\wh\O_x^2\times
\big\{ (-\pi_Q^{-1}a_1+a_2,a_1)\in {\wh F}_Q^2: a_1,a_2\in \wh\O_Q\big\}
\eea
since $a_2+\pi_Q^{-1}a_1\in \wh \O_Q$ implies that  $a_2=-\pi_Q^{-1}a_1+a_2'$, 
where $a_2':=a_2+\pi_Q^{-1}a_1\in \wh \O_X$. This proves the first two relations.

To prove the third, let $(f_2,f_1)\in \A^2(g^{~}_{I_2})\cap F^2$. Then
\be
\bc
\ord_x(f_1)\geq 0&\forall x\in X\\
\ord_x(f_2)\geq 0 &\forall x\in X\smm\{Q\}\\
\ord_x(f_2+\pi_Q^{-1}f_1)\geq 0& x=Q
\ec
\ee
The first implies that $f_1\in\F_q$, a constant function on $X$, since rational functions 
on $X$ admiting no poles are constant functions on $X$ with values in $\F_q$. In 
addition, if $f_1\not=0$, then by the third relation $\ord_Q(f_2)=-1$. This contrdicts 
with the second relation, since there is no rational function on $X$ which admits a 
simple pole in the elliptic 
curve $X$ and admits no other poles. That is to say, $f_1=0$. Using this. for $f_2$, by 
similar reason as above,  we have $f_2\in \F_q$. This verifies the third assertion in the 
lemma.

Finally, we treat th space $\A^2(g^{~}_{I_2(Q)})\cap F^2$. By the first (and the second) 
assertion(s), we have
\bea
&\A^2(g^{~}_{I_2(Q)})\cap F^2\\
=&\big(\prod_{x\in X\smm\{Q\}}\wh\O_x^2\times
\big\{ (-\pi_Q^{-1}a_1+a_2,a_1)\in {\wh F}_Q^2: a_1,a_2\in \pi_Q^{-1}\wh\O_Q\big\}
\big)\cap F^2\\
=&\left\{(f_2,f_1)\in F^2:\begin{aligned}\ord_x(f_i)\geq 0\ \ \forall x\not=Q,\\
\ord_Q(f_1)=\ord_Q(\pi_Q^{-1}a_1)\geq -1,\\
\ord_Q(f_2)=\ord_Q(\pi_Q^{-1}(-\pi_Q^{-1}a_1+a_2)\geq -2,\end{aligned}\ \ 
\exists a_1,a_2\in \wh\O_Q\right\}
\eea
Using the same argument as above, we have $f_1\in \F_q$. This implies that 
\be
a_1\in \pi_Q\wh\O_Q\qqan \ord_Q(f_2)\geq -1.
\ee
Thus, $f_2\in \F_q$, by the same argument as above. This proves the second 
equation of the forth assertion in the lemma.
In addition, $\pi_Q^{-1}(-\pi_Q^{-1}a_1+a_2)\in \wh \O_Q$.
Hence, if we set $a_1=\pi_Q a_1'$ with $a_1'\in \wh\O_Q$, then
\be
\pi_Q^{-1}\wh\O_Q\ni \pi_Q^{-1}(-\pi_Q^{-1}a_1+a_2)=\pi_Q^{-1}(-a_1'+a_1).
\ee
This proves the first equation of the forth assertion in the lemma.
\ep
With this lemma, we may continue \eqref{eq52} to obtain
\bea
H^1(F,g^{~}_{I_2})=&\frac{\frac{\big\{ (-\pi_Q^{-1}a_1+a_2,a_1)
\in {\wh F}_Q^2: a_1,a_2\in \pi_Q^{-1}\wh\O_Q\big\}}
{\big\{ (-\pi_Q^{-1}a_1+a_2,a_1)\in {\wh F}_Q^2: a_1,a_2
\in \wh\O_Q\big\}}}{\ \ \frac{\ \ \left(\prod_{x\in X\smm\{Q\}}\wh\O_x^2\times
\left\{\begin{aligned}\pi_Q^{-1}(-\pi_Q^{-1} a_1+a_2,a_1):\\
-\pi_Q^{-1} a_1+a_2=\pi_Qa_2'\\
a_1=\pi_Qa_1',\\  a_1',a_2'\in \wh\O_X\end{aligned}\right\}\right)\bigcap F^2\ \ }
{\big\{(a_2,0)\in F^2:a_2\in \F_q\big\}}\ \ }\\[1.0em]
\simeq&\frac{(-\pi_Q^{-1},1)\pi_Q^{-1}\F_q+(1,0)\pi_Q^{-1}\F_q}{\ \,  
\left(\prod_{x\in X\smm\{Q\}}\wh\O_x^2\times\left\{\begin{aligned}
\pi_Q^{-1}(-\pi_Q^{-1} a_1,a_1):\\
-\pi_Q^{-1} a_1\in \pi_Q\wh\O_X\\
a_1=\pi_Qa_1',  a_1'\in \wh\O_X\end{aligned}\right\}\right)\bigcap F^2 \ \,}\\[0.5em]
\simeq&\frac{(-\pi_Q^{-1},1)\pi_Q^{-1}\F_q+(1,0)\pi_Q^{-1}\F_q}
{(-\pi_Q^{-1},1)\pi_Q^{-1}\F_q}\simeq (\pi_Q^{-1},0)\F_q.
\eea
This then proves the first part of the following
\begin{prop}\label{l6} For the Atiyah bundle $I_3$, we have 
\begin{enumerate}
\item [(1)] The extension class of $\O_X$ by $I_2$ associated to $I_3$ is given by
\be
\kappa_{I_3,x}^{~}
=\bc (\pi_Q^{-1},0)&x=Q\\[0.6em]
\ \,(0,0)&x\not=Q.
\ec
\ee
\item[(2)] An adelic representative $g_{I_3}^{~}=(g_{I_3,x}^{~})$ for the Atiyah bundle 
$I_3$ in the quotient space
$\GL_3(F)\backslash \GL_3(\A)/\GL_3(\O)$ may be chosen as
\be
g_{I_3,x}^{~}=\bc
\bpm 1&\pi_Q^{-1}&0\\ 
0&1&\pi_Q^{-1}\\
0&0&1
\epm\qquad x=Q\\[2.0em]
\ \ \bpm 1&0&0\\ \ 0\, &\ 1\,&0\\
0&0&1 
\epm\qquad x\not=Q.
\ec
\ee
\end{enumerate}
\end{prop}
\bp
It suffices to prove the second  part. However, this is a direct consequence of the first part and Proposition\,\ref{p3}.
\ep

\subsubsection{General Cases of $I_r$}
With the cases for $I_2$ and $I_3$ treated, we are  now ready to treat $I_r$ for general $r$, for which we have the following one of our main result of this paper.

\begin{thm}\label{t7} For the Atiyah bundle $I_r$, we have 
\begin{enumerate}
\item [(1)] The extension class of $\O_X$ by $I_{r-1}$ associated to $I_r$ is given by
\be
\kappa_{I_r,x}^{~}
=\bc (\pi_Q^{-1}, \overbrace{0,\ldots,0}^{r-2})&x=Q\\[0.3em]
(\overbrace{\,0,\,0,\,\ldots,\,0\,}^{r-1})&x\not=Q.
\ec
\ee
\item[(2)] An adelic representative $g_{I_r}^{~}=(g_{I_r,x}^{~})$ for the Atiyah bundle 
$I_r$ in the quotient space
$\GL_r(F)\backslash \GL_r(\A)/\GL_r(\O)$ may be chosen as
\be
g_{I_r,x}^{~}=\bc
\bpm 1&\pi_Q^{-1}&0&\ldots&0&0\\ 
0&1&\pi_Q^{-1}&\ldots&0&0\\
0&0&1&\ldots&0&0\\
\vdots&\vdots&\vdots&\ddots&\vdots&\vdots\\
0&0&0&\ldots&1&\pi_Q^{-1}\\
0&0&0&\ldots&0&1
\epm\qquad x=Q\\[5.0em]
\qquad \bpm 1&0&0&\ldots&0&\,0\,\\ 
0&1&0&\ldots&0&0\\
0&0&1&\ldots&0&0\\
\vdots&\vdots&\vdots&\ddots&\vdots&\vdots\\
0&0&0&\ldots&1&0\\
0&0&0&\ldots&0&1
\epm\qquad x\not=Q.
\ec
\ee
\end{enumerate}
\end{thm}

\bp We prove this theorem by an induction on $r$. The cases for  $r=2,3$ 
are verified in Lemma\,\ref{cl1}, Corrary\,\ref{c5} and  Proposition\,\ref{l6}. 

Assume now that the assertions in the theorem are verified for all $I_k$ when 
$k\leq r-1$.  Since $I_r^\vee\simeq I_r$, we may equally use the exact sequence
\be\label{eq61-}
0\to I_{r-1}\lra I_r\lra \O_X\to 0 
\ee
This implies that
\be
H^1(X,\O_X^\vee\otimes I_{r-1})\simeq H^1(X, I_{r-1})\simeq H^0(X, I_{r-1}^\vee)^\vee
\simeq H^0(X, I_{r-1})^\vee=\F_q,
\ee
since $H^0(X,I_{r-1})\simeq \F_q$ by our induction assumption. Thus it is not 
surprising 
that there is one and only one non-trivial extension of $\O_X$ by $I_{r-1}$. 
To write down 
explicitly the extension class $(\kappa_{I_r,x}^{~})$ associated to the extension 
\eqref{eq61-},  similarly to the case of $I_2$ and $I_3$, we first write 
$H^1(X,I_{r-1}^{~})$ as
\bea\label{eq66}
H^1(X,I_{r-1}^{~})\simeq &(\A^{r-1}(g^{~}_{I_{r-1}(Q)})+F^{r-1})\big/
(\A^{r-1}(g^{~}_{I_{r-1}})+F^2)\\
\simeq&\frac{\A^{r-1}(g^{~}_{I_{r-1}(Q)})\big/
\A^{r-1}(g^{~}_{I_{r-1}})}{(\A^{r-1}(g^{~}_{I_{r-1}(Q)})\cap F^{r-1})
\big/(\A^{r-1}(g^{~}_{I_{r-1}})\cap F^{r-1})}
\eea

\begin{lem} We have
\bea
&\A^{r-1}(g^{~}_{I_{r-1}})\\
&=\prod_{x\in X\smm\{Q\}}\wh\O_x^{r-1}\times\left\{\begin{aligned}
\left( \sum_{i=1}^{r-1}(-\pi_Q)^{i-(r-1)}a_{i},\ldots,-\pi_Q^{-1}a_1+a_2,a_1\right)
\in {\wh F}_Q^2:\\
 a_1,a_2,\ldots, a_{r-1}\in \wh\O_Q\end{aligned}\right\}\\
&\A^{r-1}(g^{~}_{I_{r-1}(Q)})\\
&=\prod_{x\in X\smm\{Q\}}\wh\O_x^{r-1}\times\left\{\begin{aligned}
\left( \sum_{i=1}^{r-1}(-\pi_Q)^{i-(r-1)}a_{i},\ldots,-\pi_Q^{-1}a_1+a_2,a_1\right)
\in {\wh F}_Q^2:\\
 a_1,a_2,\ldots, a_{r-1}\in \pi_Q^{-1}\wh\O_Q\end{aligned}\right\}\\
&\A^{r-1}(g^{~}_{I_{r-1}})\cap F^{r-1}=\big\{(a_{r-1},\overbrace{0,\ldots,0}^{r-2})\in 
F^{r-1}:a_{r-1}\in \F_q\big\}\simeq \F_q\times\{0\}^{r-2}\\[0.5em]
&\A^{r-1}(g^{~}_{I_{r-1}(Q)})\cap F^{r-1}\\[0.5em]
&=\left\{\begin{aligned}
((-\pi_Q)^{-r+2},\ldots,-\pi_Q^{-1},1)\pi_Q^{-1}a_1\\
+((-\pi_Q)^{-r+3},\ldots,-\pi_Q^{-1},1,0)\pi_Q^{-1}a_2\\
+\ldots+(\pi_Q^{-1},1,0\ldots,0)\pi_Q^{-1}a_{r-2}\\
+(1,0\ldots,0)\pi_Q^{-1}a_{r-1}:\\
a_1,a_2,\ldots, a_{r-2},a_{r-1}\in \F_q
\end{aligned}
\right\}\bigcap F^{r-1}
\simeq\,\F_q^{r-1}
\eea
\end{lem} 
\bp
Our description of the second space $\A^{r-1}(g^{~}_{I_{r-1}(Q)})$ is a direct 
consequence of the structure for the first space 
$\A^{r-1}(g^{~}_{I_{r-1}})$, so  it suffices to calculate the later.
In addition, by our inductive assumption on $g_{I_{r-1}}^{~}$, we have
\bea
&\A^2(g^{~}_{I_2})\\
=&\prod_{x\in X\smm\{Q\}}\wh\O_x^{r-1}\times
\left\{ \begin{aligned}(a_{r-1},\ldots,a_2,a_1)\in {\wh F}_Q^{r-1}:\qquad\qquad\qquad \\
\bpm 1&\pi_Q^{-1}&0&\ldots&0&0\\ 
0&1&\pi_Q^{-1}&\ldots&0&0\\
0&0&1&\ldots&0&0\\
\vdots&\vdots&\vdots&\ddots&\vdots&\vdots\\
0&0&0&\ldots&1&\pi_Q^{-1}\\
0&0&0&\ldots&0&1
\epm
\bpm a_{r-1}\\\vdots\\ a_2\\ a_1\epm\in {\wh\O}^{r-1}_Q\end{aligned}\right\}\\
=&\prod_{x\in X\smm\{Q\}}\wh\O_x^{r-1}\times
\left\{ \begin{aligned}(a_{r-1},\ldots,a_2,a_1)\in {\wh F}_Q^{r-1}:\\
a_{r-1}+\pi_Q^{-1}a_{r-2},\ldots,a_2+\pi_Q^{-1}a_1,a_1\in {\wh \O_Q}
\end{aligned}\right\}\\
=&\prod_{x\in X\smm\{Q\}}\wh\O_x^{r-1}\times\left\{\begin{aligned}
\left( \sum_{i=1}^{r-1}(-\pi_Q)^{i-(r-1)}a_{i},\ldots,-\pi_Q^{-1}a_1+a_2,a_1\right)\in {\wh F}_Q^2:\\
 a_1,a_2,\ldots, a_{r-1}\in \wh\O_Q\end{aligned}\right\}
\eea
Indeed, if we set $a_2'=a_2+\pi_Q^{-1}a_1\in \wh\O_Q$, then 
$a_2=a_2'-\pi_Q^{-1}a_1$.
Similarly, for $3\leq j\leq r-1$, if we set $a_{j-1}'=a_{j-1}+\pi_Q^{-1}a_{j-2}\in \wh\O_Q$, 
then
\bea
a_{j-1}=&a_{j-1}'-\pi_Q^{-1}a_{j_2}=a_{j-1}'
-\pi_Q^{-1}\sum_{i=1}^{j-2}(-\pi_Q)^{i-(j-2)}a_{i}'\\
=&\sum_{i=1}^{j-1}(-\pi_Q)^{i-(j-1)}a_{i}'
\eea
This completes our proof for the first relation and hence also for the first two relations.

To prove the third, let $(f_{r-1},\ldots,f_2,f_1)\in \A^2(g^{~}_{I_2})\cap F^{r-1}$. Then
\be
\bc
\ord_x(f_1)\geq 0&\hskip -2.80cm\forall x\in X\\
\bc \ord_x(f_2)\geq 0 &\qquad\forall x\in X\smm\{Q\}\\
\ord_x(f_2+\pi_Q^{-1}f_1)\geq 0&\qquad x=Q
\ec~\\
\ldots\ldots\\
\bc
\ord_x(f_{r-1})\geq 0 &\forall x\in X\smm\{Q\}\\
\ord_x(f_{r-1}+\pi_Q^{-1}f_{r-2})\geq 0& x=Q
\ec
\ec
\ee
The first implies that $f_1\in\F_q$, a constant function on $X$, since rational functions 
on $X$ admitting no poles are constant functions on $X$ with values in $\F_q$. In 
addition, if $f_1\not=0$, then by the first in the second group relations, 
$\ord_Q(f_2)=-1$. This contradicts with the second relation of this group, since there is 
no rational function on $X$ which admits a simple pole in the elliptic curve $X$ and 
admits no other poles. That is to say, $f_1=0$. Using this, with an induction on $k+1$. 
Then if $f_k\in \F_q$ but not zero, we have $f_k$ admits a single simple pole on $X$. 
This is impossible with the same argument as above. Hence $f_k=0$ and  
$f_{k+1}\in \F_q$. Therefore, $f_1=f_2=\ldots=f_{r-2}=0$ and $f_{r-1}\in \F_q$. This 
verifies the third assertion in the lemma.

Finally, we treat th space $\A^{r-1}(g^{~}_{I_{r-1}(Q)})\cap F^{r-1}$. By the first (and 
the second) assertion(s), we have
\bea
&\A^{r-1}(g^{~}_{I_{r-1}(Q)})\cap F^{r-1}\\
=&\left(\prod_{x\in X\smm\{Q\}}\!\!\!\!\!\!\!\wh\O_x^{r-1}\!\!\times\!\!\left\{\!\!\!\!
\begin{aligned}\left( \sum_{i=1}^{r-1}(-\pi_Q)^{i-(r-1)}a_{i},\ldots,-\pi_Q^{-1}
a_1+a_2,a_1\right)\!\!\in {\wh F}_Q^2\\
 :a_1,a_2,\ldots, a_{r-1}\in \pi_Q^{-1}\wh\O_Q\end{aligned}\right\}\!\right)
 \bigcap F^{r-1}\\
=&\left\{\begin{aligned}(f_{r-1},\ldots,f_2,f_1)\in F^{r-1}:\\\
\ord_x(f_i)\geq 0\ \ \forall x\not=Q,\quad\forall i=1,\ldots,r-1\\
\exists a_1,a_2,\ldots, a_{r-1}\in \wh\O_Q\\
\ord_Q(f_1)=\ord_Q(\pi_Q^{-1}a_1)\geq -1,\\
\ord_Q(f_2)=\ord_Q(\pi_Q^{-1}(-\pi_Q^{-1}a_1+a_2)\geq -2,\\
\ldots\ldots\\
\ord_Q(f_{r-1})=\ord_Q(\pi_Q^{-1}(-\pi_Q^{-1}a_{r-2}+a_{r-1})\geq -(r-1)
\end{aligned}\right\}
\eea

Using the same argument as above, we have $\ord_Q(f_1)=0$ and hence 
$f_1\in \F_q$. This implies that
\be
a_1\in \pi_Q\wh\O_Q\qqan \ord_Q(f_2)\geq -1.
\ee
Similarly, using $f_1$ with the same reason, we get $f_2\in \F_q$. With an induction, 
the same reason implies that all $f_i\in \F_q (1\leq i\leq r-1)$. This proves the second 
equation of the forth assertion in the lemma. To prove the first equation, instead of 
using the same argument as above, we  decompose the $Q$-factor subspace 
$\left\{\begin{aligned}\left( \sum_{i=1}^{r-1}(-\pi_Q)^{i-(r-1)}a_{i},\ldots,-\pi_Q^{-1}
a_1+a_2,a_1\right)\in {\wh F}_Q^2:\\
 a_1,a_2,\ldots, a_{r-1}\in \pi_Q^{-1}\wh\O_Q\end{aligned}\right\}$ of 
 $\A^{r-1}(g^{~}_{I_{r-1}(Q)})$ as
\bea
\left\{\begin{aligned}
((-\pi_Q)^{-r+2},\ldots,-\pi_Q^{-1},1)\pi_Q^{-1}a_1+((-\pi_Q)^{-r+3},\ldots,-\pi_Q^{-1},
1,0)\pi_Q^{-1}a_2\\
+\ldots+(\pi_Q^{-1},1,0\ldots,0)\pi_Q^{-1}a_{r-2}
+(1,0\ldots,0)\pi_Q^{-1}a_{r-1}:\\
a_1,a_2,\ldots, a_{r-2},a_{r-1}\in \wh\O_Q
\end{aligned}
\right\}
\eea
Essentially with the same argument, by looking the last component $\pi_Q^{-1}a_1$ of 
$(-\pi_Q)^{-r+2},\ldots,-\pi_Q^{-1},1)\pi_Q^{-1}a_1$, to have an element $(f_{r-1},\ldots, 
f_2,f_1)$ of  $F^{r-1}$ in the intersection, by working with poles and zeros of rational 
functions over elliptic curves, we conclude that $a_1\in \pi_Q\wh\O_X$. Based on this, 
if we set $a_1=\pi_Qa_1'$, then, by looking at the part 
\bes
((-\pi_Q)^{-r+2},\ldots,-\pi_Q^{-1},0)a_1'+((-\pi_Q)^{-r+3},\ldots,
-\pi_Q^{-1},1,0)\pi_Q^{-1}a_2 
\ees
for $f_2$, we conclude that $\pi_Q^{-1}a_2+a_1'\in \wh\O_Q$. This implies that 
$a_2=\pi_Qa_2'\in \pi_Q\wh\O_Q$.  Thus inductively, for $a_k=\pi_Qa_k'\in 
\pi_Q\wh\O_Q$, we have for the element
\bes
((-\pi_Q)^{-r+2},\ldots,-\pi_Q^{-k},0\ldots,0)a_k'+((-\pi_Q)^{-r+3},\ldots,-\pi_Q^{-k+1},
0\ldots,0)\pi_Q^{-1}a_{k+1} 
\ees
associated to $f_{k+1}$, we conclude that 
\be
a_{k+1}=\pi_Qa_{k+1}'\in \pi_Q\wh\O_Q\qquad\forall 1\leq i\leq r-1.
\ee
This proves the first equation of the forth assertion in the lemma.
\ep
With this lemma, we now continue the calculation in \eqref{eq66} to obtain
\bea\label{eq77}
&H^1(X,I_{r-1}^{~})\\
\simeq &
\frac{\ \, 
\left\{\begin{aligned}((-\pi_Q)^{-r+2},\ldots,-\pi_Q^{-1},1)a_1\\
+((-\pi_Q)^{-r+3},\ldots,-\pi_Q^{-1},1,0)a_2\\
+\ldots+(\pi_Q^{-1},1,0\ldots,0)a_{r-2}\\
+(1,0\ldots,0)
a_{r-1}:\\a_1,a_2,\ldots,a_{r-1}\in \pi_Q^{-1}\wh\O_Q\end{aligned}\right\}
\Big/ \left\{\begin{aligned}((-\pi_Q)^{-r+2},\ldots,-\pi_Q^{-1},1)a_1\\
+((-\pi_Q)^{-r+3},\ldots,-\pi_Q^{-1},1,0)a_2\\
+\ldots+(\pi_Q^{-1},1,0\ldots,0)a_{r-2}\\
+(1,0\ldots,0)
a_{r-1}:\\a_1,a_2,\ldots,a_{r-1}\in \wh\O_Q
\end{aligned}
\right\}\ \ \,}
{\ \,  \left\{\begin{aligned}((-\pi_Q)^{-r+2},\ldots,-\pi_Q^{-1},1)a_1\\
+((-\pi_Q)^{-r+3},\ldots,-\pi_Q^{-1},1,0)a_2\\
+\ldots+(\pi_Q^{-1},1,0\ldots,0)a_{r-2}\\
+(1,0\ldots,0)
a_{r-1}:\\a_1,a_2,\ldots,a_{r-1}\in\pi_Q^{-1} \F_q\end{aligned}\right\}  
 \Big/(\F_q\times\{0\}^{r-2})}\\
 \simeq&
 \frac{\ \, 
\left(
\begin{aligned}
((-\pi_Q)^{-r+2},\ldots,-\pi_Q^{-1},1)\pi_Q^{-1}\F_q\\
+((-\pi_Q)^{-r+3},\ldots,-\pi_Q^{-1},1,0)\pi_Q^{-1}\F_q\\
+\ldots+(\pi_Q^{-1},1,0\ldots,0)\pi_Q^{-1}\F_q\\
+(1,0\ldots,0)\pi_Q^{-1}\F_q
\end{aligned}
\right)
\ \ \,}
{\ \,  
\left(
\begin{aligned}
((-\pi_Q)^{-r+2},\ldots,-\pi_Q^{-1},1)\pi_Q^{-1}\F_q\\
+((-\pi_Q)^{-r+3},\ldots,-\pi_Q^{-1},1,0)\pi_Q^{-1}\F_q\\
+\ldots+(\pi_Q^{-1},1,0\ldots,0)\pi_Q^{-1}\F_q
\end{aligned}
\right)  
}
\simeq (\pi_Q^{-1}\F_q,\overbrace{0\ldots,0}^{r-2})
\eea
Therefore, $\kappa_{I_{r-1}^{~}}^{~}$ is given as in the theorem.

This completes the proof of the Theorem.
\ep

\section{Elements in $H^0(F,g_{I_r}^{~}(mQ))$}

\subsection{Loca Conditions}
To start with, we here give a very important property for global sections in 
$H^0(F,g_{I_r}^{~}(mQ))$.

Since $g_{I_r}^{~}(mQ)$ is semi-stable of degree $mr$, by the Riemann-Roch 
theorem and the vanishing theorem (for semi-stable bundles, we have
\be
\dim_{\F_q} H^0(F,g_{I_r}^{~}(mQ))=mr.
\ee
Let ${\bf f}=(f_r,\ldots,f_1)\in H^0(F,g_{I_r}^{~}(mQ))$. Then, by definition, we have 
the following characterizing condition 
\be
(g_{I_r}^{~}(mQ)){\bf f}^t\in \O^r.
\ee
Hence by Theorem\,\ref{t7}, $(f_r,\ldots,f_1)$ satisfies the conditions
\be
\bc
\bpm 1&\pi_Q^{-1}&0&\ldots&0&0\\ 
0&1&\pi_Q^{-1}&\ldots&0&0\\
0&0&1&\ldots&0&0\\
\vdots&\vdots&\vdots&\ddots&\vdots&\vdots\\
0&0&0&\ldots&1&\pi_Q^{-1}\\
0&0&0&\ldots&0&1
\epm\bpm f_r\\ f_{r-1}\\\vdots\\f_1\epm\in \pi_Q^{-m}\wh \O_Q^r&x=Q\\[5.0em]
\ \,\bpm 1&0&0&\ldots&0&\,0\,\\ 
0&1&0&\ldots&0&0\\
0&0&1&\ldots&0&0\\
\vdots&\vdots&\vdots&\ddots&\vdots&\vdots\\
0&0&0&\ldots&1&0\\
0&0&0&\ldots&0&1
\epm\bpm \bpm f_r\\ f_{r-1}\\\vdots\\ f_1\epm\epm\in \wh O_x^r&x\not=Q
\ec
\ee
That is to say, for $x\not=Q$, $f_j\in \wh \O_x\ \forall \ 1\leq j\leq r$. And, at the point 
$Q$, we have
\be\label{eq68}
\bc
f_r&= -\pi_Q^{-1}f_{r-1}+\pi_Q^{-m}f_r'\quad \in -\pi_Q^{-1}f_{r-1}+
\pi_Q^{-m}\wh \O_Q\\
f_{r-1}&=-\pi_Q^{-1}f_{r-2}+ \pi_Q^{-m}f_{r-1}'\in -\pi_Q^{-1}f_{r-2}+ 
\pi_Q^{-m}\wh \O_Q\\
\ldots&\ldots\ldots\\
f_{3}&=-\pi_Q^{-1}(-\pi_Q^{-1} \pi_Q^{-m}f_1'\ \ \ + \pi_Q^{-m}f_2')\ \ \ + \pi_Q^{-m}f_3'\\
=&\pi_Q^{-m}((-\pi_Q)^{-2}f_1'+(-\pi_Q^{-1})f_2'\\
f_{2}&=-\pi_Q^{-1}f_1\ \ \ + \pi_Q^{-m}f_2'\quad \in -\pi_Q^{-1}f_1\ \ +
 \pi_Q^{-m}\wh \O_Q\\
f_1&=\hskip 2.0cm   \pi_Q^{-m}f_1'\quad \in\hskip 1.9cm   \pi_Q^{-m}\wh \O_Q
\ec
\ee
for some $f_r',\ldots, f_2', f_1'\in \wh\O_Q$.
We obtain
\be\label{eq68}
\bc
f_r&=\pi_Q^{-m}\sum_{i=1}^{r-1}( -\pi_Q)^{-r+i}f_i'\\
f_{r-1}&=\pi_Q^{-m}\sum_{i=1}^{r-2}( -\pi_Q)^{-r-1+i}f_i'\\
\ldots&\ldots\ldots\\
f_{2}&=\pi_Q^{-m}(-\pi_Q^{-1}f_1'+ f_2')\\
f_1&= \pi_Q^{-m}\qquad \ \  \,f_1'
\ec
\ee
Consequently, 
\bea
&(f_r,f_{r-1}\ldots,f_2,f_1)\\
=&\pi_Q^{-m}\left(\sum_{i=1}^{r-1}( -\pi_Q)^{-r+i}f_i',\ \sum_{i=1}^{r-2}( -\pi_Q)^{-r-1+i}f_i',\ldots, -\pi_Q^{-1}f_1'+ f_2',f_1'\right)\\
=&\pi_Q^{-m}
\left(
\begin{aligned}
\left((-\pi_Q)^{-r+1},(-\pi_Q)^{-r+2},\ldots,-\pi_Q^{-1},1\right)f_1'\\
+\left(-\pi_Q)^{-r+2},(-\pi_Q)^{-r+3},\ldots,1,0\right)f_2'+\ldots\\
+\left(-\pi_Q,1,\ldots,0,0\right)f_{r-1}'+\left(1,0,\ldots,0\right)f_r'
\end{aligned}
\right)
\eea
for some $f_r',f_{r-1}',\ldots, f_2', f_1'\in \wh\O_Q$.

\subsection{Cases $I_2$ and $I_3$}

First we consider case for $I_2$.

When $m=-1$, we have, at $Q$,
\be
(f_2,f_1)=\pi_Q^{-1}\big(\,(-\pi_Q^{-1},1)f_1'+(1,0)f_2'\,\big)\qquad\exists f_1', f_2'
\in \wh\O_Q
\ee
This implies that $f_1'=\pi_Qf_1''\in \pi_Q\wh\O_Q$ since otherwise $f_1$ is regular at 
all $x\not=Q$ and admits a simple pole at $Q$. This is impossible, since $X$ is an 
elliptic curve. Therefore, 
\be
(f_2,f_1)
=\big(\,(-\pi_Q^{-1},1)f_1''+(\pi_Q^{-1},0)f_2'\,\big)=(\pi_Q^{-1}(f_2'-f_1''), 
f_1'')=(f_2'',f_1'')
\ee
for some $f_1'',  f_2''\in \wh\O_Q$, since similarly, $f_2'-f_1''\in\pi_Q \O_Q$. All these 
then complete a proof of  the following

\begin{lem} If $(f_1,f_2)\in H^0(F,g_{I_2(Q)}^{~})$, we have 
\be
f_1,f_2\in \F_q\qqan
(f_2,f_1)=(f_2'',f_1'')\quad {\mathrm{at}}\ Q
\ee
for some $f_1'',  f_2''\in \F_q\subset\wh\O_Q$ defined above.
\end{lem}

Next, we treat the case $m=2$ for $I_2(2Q)$. This time, we go back to the condition
\be
f_1\in \pi_Q^{-2}\wh\O_Q\qqan f_2+\pi^{-1}_Qf_1\in \pi_Q^{-2}\wh\O_Q
\ee
From the first, we see that $\ord_Q(f_1)=0,-2$. Moreover, 
\begin{enumerate}
\item [(a)] If $\ord_Q(f_1)=0$, $\ord_Q(f_2)=0,-2$. This gives a three (=one+two) 
dimensional subspace in $H^0(F,g_{I_2(2Q)}^{~})$
\item [(b)] If $\ord_Q(f_1)=-2$, $\ord_Q(f_2)=-3$. This gives an one dimensional 
subspace in $H^0(F,g_{I_2(2Q)}^{~})$.
\end{enumerate}

Now we are ready to study $H^0(F,g_{I_2(mQ)}^{~})$ for general $m$ by an induction 
on $m$. Assume that when $m=k$, there is a basis of $H^0(F,g_{I_2(kQ)}^{~})$
constructed from the table

\begin{table}[htp]
\centering
\begin{tabular}{|c||c|c||c|c|c|c|}\hline
$\ord_Q(f_1)\backslash \ord_Q(f_2)$&0&-2&$\ldots$&-k+1&-k&-k-1\\\hline\hline
0&O&O&$\ldots$&O&O&X\\\hline
-2&O&O&$\ldots$&O&O&X\\\hline
$\vdots$&$\vdots$&$\vdots$&$\ddots$&$\vdots$&$\vdots$&$\vdots$\\\hline
-k+1&O&O&$\ldots$&O&O&X\\\hline
-k&X&X&$\ldots$&X&X&O\\\hline
\end{tabular}
\caption{Pole orders occurring for $(f_2,f_1)\in H^0(F,g_{I_2(kQ)}^{~})$}
\label{table:1}
\end{table}

\noindent
induced from the relations 
\be
f_1\in \pi_Q^{-k}\wh\O_Q\qqan 
f_2+\pi_Q^{-1}f_1\in \pi_Q^{-k}\wh\O_Q.
\ee
Here in the table above, O (resp. X), means that the values for $(\ord_Q(f_1), 
\ord_Q(f_2))$ occurs (resp. does not occur). In particular, the total dimension 
$H^0(F,g_{I_2(kQ)}^{~})$ is given by $(k-1)+k+1=2k$.
Then for $(f_2,f_1)\in H^0(F,g_{I_2((k+1)Q)}^{~})$, with the same discussion, we see 
that
\bea
\ord_Q(f_1)=&0,-2,-3,\ldots,-(k+1)\\
\ord_Q(f_2)=&0,-2,-3,\ldots,-(k+1),-(k+1)-1.
\eea
Moreover, from Table\,\ref{table:1}, not only we should enlarge the table, but to 
recheck the cases involved. It is not difficult to deduces the following
\begin{table}[htp]
\centering
\begin{tabular}{|c||c|c||c|c|c|c|c|}\hline
$\ord_Q(f_1)\backslash \ord_Q(f_2)$&0&-2&$\ldots$&-k+1&-k&-(k+1)&-(k+1)-1\\
\hline\hline
0&O&O&$\ldots$&O&O&O&X\\\hline
-2&O&O&$\ldots$&O&O&O&X\\\hline
$\vdots$&$\vdots$&$\vdots$&$\ddots$&$\vdots$&$\vdots$&$\vdots$&$\vdots$\\\hline
-k+1&O&O&$\ldots$&O&O&O&X\\\hline
-k&O&O&$\ldots$&O&O&O&X\\\hline
-(k+1)&X&X&$\ldots$&X&X&X&O\\\hline
\end{tabular}
\caption{Pole orders occurring for $(f_2,f_1)\in H^0(F,g_{I_2((k+1)Q)}^{~})$}
\label{table:2}
\end{table}

\noindent
In particular, the total dimension $H^0(F,g_{I_2((k+1)Q)}^{~})$ is given by $k+
(k+1)+1=2(k+1).$

With $I_2$ done, next we check $I_3$. The difference is that one more relation is 
added, namely
\be
f_3+\pi_Q^{-1}f_2\in \pi_Q^{-m}\wh\O_Q.
\ee
For this, with carefully case-by-case checking, we arrive at the following 
\begin{table}[htp]
\centering
\begin{tabular}{|c||c|c||c|c|c|c|c|}\hline
$\ord_Q(f_2)\backslash \ord_Q(f_3)$&0&-2&$\ldots$&-m+1&-m&-m-1&-m-2\\
\hline\hline
0&O&O&$\ldots$&O&O&X&X\\\hline
-2&O&O&$\ldots$&O&O&X&X\\\hline
$\vdots$&$\vdots$&$\vdots$&$\ddots$&$\vdots$&$\vdots$&$\vdots$&$\vdots$\\\hline
-m+1&O&O&$\ldots$&O&O&X&X\\\hline
-m&X&X&$\ldots$&X&X&O&X\\\hline
-m-1&X&X&$\ldots$&X&X&X&O\\\hline
\end{tabular}
\caption{Pole orders occurring for $(f_3,f_2)$ of $(f_3,f_2,f_1)\in 
H^0(F,g_{I_3(mQ)}^{~})$}
\label{table:3}
\end{table}

\noindent
In particular, this implies that the dimension of $H^0(F,g_{I_3(mQ)}^{~})$ is given by
\be
(m-1)+(m-1)+m+1+1=3m.
\ee

\subsection{Case for $I_r$}

Continuing the discussion in the previous subsubsection, in the case for $I_4$, one 
more condition should be  added, namely
\be
f_4+\pi_Q^{-1}f_3\in \pi_Q^{-m}\wh\O_Q.
\ee
For this, with carefully case-by-case checking, we arrive at the following 
\begin{table}[h!]
\centering
\begin{tabular}{|c||c|c||c|c|c|c|c|c|}\hline
$\ord_Q(f_3)\backslash\ord_Q(f_4)$&0&-2&$\ldots$&-m+1&-m&-m-1&-m-2&-m-3
\\\hline\hline
0&O&O&$\ldots$&O&O&X&X&X\\\hline
-2&O&O&$\ldots$&O&O&X&X&X\\\hline
$\vdots$&$\vdots$&$\vdots$&$\ddots$&$\vdots$&$\vdots$&$\vdots$&$\vdots$\\\hline
-m+1&O&O&$\ldots$&O&O&X&X&X\\\hline
-m&X&X&$\ldots$&X&X&O&X&X\\\hline
-m-1&X&X&$\ldots$&X&X&X&O&X\\\hline
-m-2&X&X&$\ldots$&X&X&X&X&O\\\hline
\end{tabular}
\caption{Pole orders occurring for $(f_4,f_3)$ of $(f_4,\ldots,f_1)\in 
H^0(F,g_{I_4(mQ)}^{~})$}
\label{table:4}
\end{table}

\noindent
In particular, this implies that the dimension of $H^0(F,g_{I_4(mQ)}^{~})$ is given by
\be
(m-1)+(m-1)+(m-1)+m+1+1+1=4m.
\ee

Therefore, for general $I_r$, we should have
\begin{table}[h!]
\centering
\begin{tabular}{|c||c|c||c|c|c|c|c|c|c|}\hline
$o(f_{r-1})\backslash o(f_r)$&0&-2&$\ldots$&-m+1&-m&-m-1&\ldots&-m-(r-2)&-m-
(r-1)\\\hline\hline
0&O&O&$\ldots$&O&O&X&$\ldots$&X&X\\\hline
-2&O&O&$\ldots$&O&O&X&$\ldots$&X&X\\\hline
$\vdots$&$\vdots$&$\vdots$&$\ddots$&$\vdots$&$\vdots$&$\vdots$&$\ddots$&$\vdots$&$\vdots$\\\hline
-m+1&O&O&$\ldots$&O&O&X&$\ldots$&X&X\\\hline
-m&X&X&$\ldots$&X&X&O&$\ldots$&X&X\\\hline
-m-1&X&X&$\ldots$&X&X&X&$\ldots$&X&X\\\hline
-m-2&X&X&$\ldots$&X&X&X&$\ldots$&X&X\\\hline
$\vdots$&$\vdots$&$\vdots$&$\ddots$&$\vdots$&$\vdots$&$\vdots$&$\ddots$&$
\vdots$&$\vdots$\\\hline
-m-(r-3)&X&X&$\ldots$&X&X&X&$\ldots$&O&X\\\hline
-m-(r-2)&X&X&$\ldots$&X&X&X&$\ldots$&X&O\\\hline
\end{tabular}
\caption{Pole orders occurring for $(f_r,f_{r-1})$ of $(f_r,\ldots,f_1)\in H^0(F,g_{I_r(mQ)}^{~})$}
\label{table:5}
\end{table}

\noindent
Here $o(f):=\ord_Q(f)$.
In particular, this implies that the dimension of $H^0(F,g_{I_r(mQ)}^{~})$ is given by
\be
\overbrace{(m-1)+\ldots+(m-1)}^{r-1}+m+\overbrace{1+\ldots+1}^{r-1}=rm.
\ee

\section{Rank $r$ Codes $C_F(D,g_{I_r}^{~}(mQ))$}
\subsection{Some General Results}\label{s5.1}

Let $p_1,p_2,\ldots, p_n$ be mutually distinct $\F_q$-rational points of $X$. Set 
$D=p_1+p_2+\ldots,+p_n$ be the associated  divisor on $X$. The, for any $m\geq 1$, 
using the constructions in  \cite{We}, we obtain a rank $r$ code 
$C_F(D,g_{I_2}^{~}(mQ))$, defined as the linear space 
\be
\left\{\begin{aligned}\big(f_2(p_1), f_1(p_1);f_2(p_2), f_1(p_2);\ldots; 
f_2(p_n), f_1(p_n) \big): \\(f_2, f_1)\in H^0(F,g_{I_2}^{~}(mQ))\hskip 2.0cm 
\end{aligned}\right\}
\ee
By Table\,\ref{table:1}, $\ord_Q(f_1)\geq -m$, $\ord_Q(f_2)\geq -m-1$ and it is 
possible to have
\be
\ord_Q(f_1)= -m\qqan\ord_Q(f_2)= -m-1
\ee

Now we are ready to state the next main theorem of this paper.

\begin{prop}\label{p9} Let $(n,m)\in \Z_{\geq 0}^2$.
Assume that $p_1,\ldots,p_n,\, Q$ are $\F_q$-rational points on the elliptic curve $X$  
with additive operation $\oplus$ satisfy the following conditions.
\begin{enumerate}
\item [(1)] $Q$ is the zero element of the group $(X(\F_q),\oplus)$,
\item [(2)] $p_1\oplus\ldots\oplus p_{m}\oplus p_{m+1}=Q$
\item [(3)] $p_{m}\oplus p_{m+1}\in \{p_{m+2},\ldots, p_n\}$  
\end{enumerate}
Set $D=p_1+p_2+\ldots+p_n$ be the divisor on $X$ associated to the $p_i$'s.
Then for the $D$-balanced, semi-stable $g_{I_2}^{~}(mQ)\in\GL_2(\A)$, the 
dimension and the minimal distance of  the rank $r$ code space $\CFm$ are given by
\be
k_{D, g_{I_2}^{~}(mQ)}=2m\qqan d_{D, g_{I_2}^{~}(mQ)}=2(n-m)-1,
\ee
respectively. In particular, we have
\be
k_{D, g_{I_2}^{~}(mQ)}+d_{D, g_{I_2}^{~}(mQ)}=2n-1=\ell_{D, g_{I_2}^{~}(mQ)}-1.
\ee
\end{prop}

\bp We begin with the following well-known

\begin{lem}\label{le5} Let $\sum_{i=1}^sn_iP_i$ be a divisor on  the elliptic curve 
$X$ with $Q$ as its zero element. Then  
\be
\sum_{i=1}^sn_iP_i=(f)
\ee
for a certain rational function $f$ if and only if
\begin{enumerate}
\item [(1)] $\sum_{i=1}^sn_i\deg(P_i)=0$, and
\item[(2)] $\bigoplus_{i=1}^s [n_i]P_i=Q$.
\end{enumerate}
Here, for a closed point $P\in X$ and an integer $n$, we write $[n]P$ for $\overbrace{P\oplus\ldots\oplus P}^n$.
\end{lem}

Consequently, by the condition (1) and (3) in the theorem, it is possible to choose an 
element  $f_1\in F$ such that
\be
(f_{1,0})=p_1+p_2+\ldots+(p_{m}\oplus p_{m+1})-mQ.
\ee
Similar, by the condition (1) and (2) in the theorem, it is possible to choose an 
element  $f_{2}\in F$ such that
\be
(f_{2,0})=p_1+p_2+\ldots+p_{m}+p_{m+1}-(m+1)Q.
\ee
Obviously,
\be
(f_{2,0} f_{1,0})\in H^0(F,g_{I_r}^{~}(mQ)).
\ee
Therefore,
\be
\sum_{i,j=1}^{n, r}\delta_{\ord_{p_i}(f_j)\geq 1}=m+(m+1)=2m+1.
\ee
On the other hand, by Table\,\ref{table:1}, we have
\bea
&\max\Big\{\sum_{i,j=1}^{n, 2}\delta_{\ord_{p_i}(f_j)\geq 1}:
(f_2,f_1)\in H^0(F,g_{I_2}^{~}(mQ))\Big\}\leq m+(m+1)=2m+1
\eea
Therefore,
\be
\max\Big\{\sum_{i,j=1}^{n, 2}\delta_{\ord_{p_i}(f_j)\geq 1}:
(f_2,f_1)\in H^0(F,g_{I_2}^{~}(mQ))\Big\}=2m+1.
\ee
Hence, by the fact that the length of the code $\CFm$ is $nr$, from Lemma\,17 of 
\cite{We}, we conclude that  the minimal distance of $\CFm$ is given by
\bea\label{eq91}
d_{D,g_{I_2}^{~}(mQ)}=&\,2n-\max\Big\{\sum_{i,j=1}^{n, 2}
\delta_{\ord_{p_i}(f_j)\geq 1}:(f_2,f_1)\in H^0(F,g_{I_2}^{~}(mQ))\Big\}\\
=&\,2n-(2m+1)=2(n-m)-1
\eea

Next, we calculate the dimension $k_{D,g_{I_2}^{~}(mQ)}$ of our rank $2$ codes. 
Note that $\deg(g_{I_2}^{~}(mQ-D))=2(m-n)<0$, by the vanishing theorem,
\be
H^0(F,g_{I_2}^{~}(mQ-D))=\{0\}.
\ee
This implies that
\be\label{eq93}
k_{D,g_{I_2}^{~}(mQ)}=h^0(F,g_{I_2}^{~}(mQ))-h^0(F,g_{I_2}^{~}(mQ-D))=2m-0=2m,
\ee
Therefore, by \eqref{eq91} and \eqref{eq93},
\be
k_{D,g_{I_2}^{~}(mQ)}+d_{D,g_{I_2}^{~}(mQ)}=(2(n-m)-1)+2m=2n-1.
\ee
This completes the our proof.
\ep

With similar arguments, we have the following
\begin{prop}\label{p10} Let $(n,m)\in \Z_{\geq 0}^2$.
Assume that $p_1,\ldots,p_n,\, Q$ are $\F_q$-rational points on the elliptic curve $X$  
with additive operation $\oplus$ satisfy the following conditions.
\begin{enumerate}
\item [(1)] $Q$ is the zero element of the group $(X(\F_q),\oplus)$,
\item [(2)] $p_1\oplus\ldots\oplus p_{m}\oplus p_{m+1}\oplus p_{m+2}=Q$
\item [(3)] $p_{m}\oplus p_{m+1}\oplus p_{m+2}, p_{m+1}\oplus 
p_{m+2}\in \{p_{m+3},\ldots, p_n\}$  
\end{enumerate}
Set $D=p_1+p_2+\ldots+p_n$ be the divisor on $X$ associated to the $p_i$'s.
Then for the $D$-balanced, semi-stable $g_{I_3}^{~}(mQ)\in\GL_3(\A)$, the 
dimension and the minimal distance of  the rank $3$ code space $\CFm$ are given by
\be
k_{D, g_{I_3}^{~}(mQ)}=3m\qqan d_{D, g_{I_3}^{~}(mQ)}=3(n-m)-3,
\ee
respectively. In particular, we have
\be
k_{D, g_{I_3}^{~}(mQ)}+d_{D, g_{I_3}^{~}(mQ)}=3n-3=\ell_{D, g_{I_2}^{~}(mQ)}-3.
\ee
\end{prop}

More  generally, we have the following 
\begin{thm}\label{t9} Let $(n,r,m)\in \Z_{\geq 0}^3$ satisfying $n\geq m+2r-2$.
Assume that $p_1,\ldots,p_n,\, Q$ are $\F_q$-rational points on the elliptic curve $X$  
with additive operation $\oplus$ satisfy the following conditions.
\begin{enumerate}
\item [(1)] $Q$ is the zero element of the group $(X(\F_q),\oplus)$,
\item [(2)] $p_1\oplus\ldots\oplus p_{m}\oplus p_{m+1}\oplus \ldots\oplus 
p_{m+r-1}=Q$
\item [(3)] $p_{m}\oplus p_{m+1}\oplus\ldots\oplus  p_{m+r-2}\oplus p_{m+r-1},
 \ldots, p_{m+r-2}\oplus p_{m+r-1}\in \{p_{m+r},\ldots, p_n\}$  
\end{enumerate}
Set $D=p_1+p_2+\ldots+p_n$ be the divisor on $X$ associated to the $p_i$'s.
Then for the $D$-balanced, semi-stable $g_{I_r}^{~}(mQ)\in\GL_r(\A)$, the 
dimension and the minimal distance of  the rank $r$ code space $\CFm$ are given by
\be
k_{D, g_{I_r}^{~}(mQ)}=rm\qqan d_{D, g_{I_r}^{~}(mQ)}=r(n-m)-\frac{r(r-1)}{2},
\ee
respectively. In particular, we have
\be
k_{D, g_{I_r}^{~}(mQ)}+d_{D, g_{I_r}^{~}(mQ)}=rn-\frac{r(r-1)}{2}
=\ell_{D, g_{I_2}^{~}(mQ)}-\frac{r(r-1)}{2}.
\ee
\end{thm}

\subsection{MDS Codes $C_F(D,g_{I_2}^{~}(mQ))$}

The previous subsection gives some general results for $C_F(D,g_{I_r}^{~}(mQ))$.
It is then a natural question when $C_F(D,g_{I_r}^{~}(mQ))$ becomes MDS. To 
simplify our discussions, we assume $r=2$. 

By definition, for  MDS  codes $C_F(D,g_{I_2}^{~}(mQ))$,
\be
k_{D, g_{I_2}^{~}(mQ)}+d_{D, g_{I_2}^{~}(mQ)}=\ell_{D, g_{I_2}^{~}(mQ)}+1.
\ee
On the other hand, by the discussion in \S\ref{s5.1}, 
\be
k_{D, g_{I_2}^{~}(mQ)}=2m\qqan \ell_{D, g_{I_2}^{~}(mQ)}=2n.
\ee
Hence, for MDS codes,
\be
d_{D, g_{I_r}^{~}(mQ)}=2(n-m)+1.
\ee
Recall that, by Lemma 17 of \cite{We},
\bea\label{eq91}
d_{D,g_{I_2}^{~}(mQ)}=&\,2n-\max\Big\{\sum_{i,j=1}^{n, 2}
\delta_{\ord_{p_i}(f_j)\geq 1}:(f_2,f_1)\in H^0(F,g_{I_2}^{~}(mQ))\Big\}
\eea
All these imply that, for MDS codes  $C_F(D,g_{I_2}^{~}(mQ))$,
\be
\max\Big\{\sum_{i,j=1}^{n, 2}
\delta_{\ord_{p_i}(f_j)\geq 1}:(f_2,f_1)\in H^0(F,g_{I_2}^{~}(mQ))\Big\}=2m-1.
\ee
Our aim in this subsection is to find a global  section 
$(f_2,f_1)\in H^0(F,g_{I_2}^{~}(mQ))$ such that
\be\label{eq117}
\sum_{i=1}^{n}
\delta_{\ord_{p_i}\geq 1}(f_1)=m-1\qqan \sum_{i=1}^{n}
\delta_{\ord_{p_i}\geq 1}(f_2)=m.
\ee

Recall that, from the discussion in \S\ref{s5.1},
for  $(f_2,f_1)\in H^0(F,g_{I_2(mQ)}^{~})$,
\be
\ord_Q(f_1)\geq -m\qqan \ord_Q(f_2)\geq -(m+1).
\ee
and there is only one $\F_q$-subspace $(f_{2,0},f_{1,0})\F_q$ of 
$H^0(F,g_{I_2(mQ)}^{~})$ such that
\be
\ord_Q(f_1)= -m\qqan \ord_Q(f_2)= -(m+1).
\ee

For example, if the following conditions are satisfied, the corresponding codes 
$C_F(D,g_{I_2}^{~}(mQ))$ is MDS.

\begin{enumerate}
\item [(0)] $Q$ is the zero moment of the elliptic curve $X/\F_q$,
\item [(1)] $(f_1)=p_1+\ldots +p_{m-2} +2p_{m-1}-mQ$,
\item [(2)] $(f_2)=p_1+\ldots+p_{m-1}+2p_{m}-(m+1)Q$,
\item [(3)] $p_{m-1}=[2]p_{m}$,
\item [(4)] $p_1\oplus p_1\oplus\ldots\oplus p_{m-2}\oplus [2]p_{m-1}
=p_1\oplus p_1\oplus\ldots\oplus p_{m-2}\oplus p_{m-1}\oplus [2]p_{m}=Q$.
\end{enumerate}

Obviously, if (3) and (4) are satisfied, then so is (1) and (2) by lemma\,\ref{le5}.
For example, if $X(F_q)$ contains a cyclic subgroup of order 4 generated by $p_m$, 
then (3) and (4) are satisfied by taking $p_1,\ldots, p_{m-2}$ satisfying
\be
p_1\oplus\ldots \oplus p_{m-2}=Q.
\ee
Motivated by this, we may also take any cyclic factor of $X/\F_q$ generated by 
$p_m$, then assume that $p_m, [\a_1]p_m=p_{m-1}, [\a_2]p_{m_1}=p_{m-2}.\ldots$ 
to obtain rank $r$ MDS codes $C_F(D,g_{I_r}^{~}(mQ))$.
In any cases, there are many many ways to obtain MDS codes using 
$C_F(D,g_{I_r}^{~}(mQ))$.

~\vskip 14.60cm

L. WENG

Faculty of Mathematics 

Kyushu University

Fukuoka 819-0395

JAPAN

E-Mail: weng@math.kyushu-u.ac.jp


\begin{thebibliography}{80}

\bibitem{A} M.F. Atiyah,  Vector bundles over an elliptic curve. 
Proc. London Math. Soc. (3) 7 (1957) 414-452. 

\bibitem{H} R. Hartshorne, {\it Algebraic geometry}. GTM 52. Springer, 1977. 

\bibitem{Mu} D. Mumford, J. Fogarty, F. Kirwan, {\it  Geometric invariant theory}. Third 
edition. Ergebnisse der Mathematik und ihrer Grenzgebiete (2), 34. Springer, 1994. 
xiv+292 pp.

\bibitem{S} H. Stichtenoth, {\it Algebraic Function Fields and Codes}, GTM 254, 
Springer, 2009. xiv+355 pp.

\bibitem{T} L.W. Tu, Semi-stable bundles over an elliptic curve, Advance in Math. 98, 
(1993) 1-26. 

\bibitem{We} L. Weng, Codes and Stability, arXiv:1806.04319
\end{thebibliography}
\end{document}